\documentstyle[12pt]{article}
\newcommand{\sect}[1]{\setcounter{equation}{0}\section{#1}\indent}
\renewcommand{\theequation}{\thesection.\arabic{equation}}
\renewcommand{\thefootnote}{\fnsymbol{footnote}}
\newcommand{\EQ}{\begin{equation}}
\newcommand{\EN}{\end{equation}}
\newcommand{\bea}{\begin{eqnarray}}
\newcommand{\ena}{\end{eqnarray}}
\newcommand{\vs}[1]{\vspace{#1 mm}}

\newcommand{\e}{\epsilon}

\newcommand{\G}{\Gamma}

\newcommand{\pa}{\partial}

\newcommand{\uda}{\nearrow \kern-1em \searrow}

\newcommand{\CMP}[1]{Comm.\ Math.\ Phys.\ {\bf #1}}
\newcommand{\PR}[1]{Phys.\ Rev.\ {\bf #1}}
\newcommand{\PRL}[1]{Phys.\ Rev.\ Lett.\ {\bf #1}}
\newcommand{\PTP}[1]{Prog.\ Theor.\ Phys.\ {\bf #1}}

\newcommand{\AJ}[1]{Astorophys. \ J.\ {\bf #1}}
\newcommand{\JMP}[1]{J.\ Math.\ Phys.\ {\bf #1}}

\newcommand{\la}{\lambda}

\makeatletter
\def\eqnarray{%
 \stepcounter{equation}%
 \let\@currentlabel=\theequation
 \global\@eqnswtrue
 \global\@eqcnt\z@
 \tabskip\@centering
 \let\\=\@eqncr
 $$\halign to \displaywidth\bgroup\@eqnsel\hskip\@centering
 $\displaystyle\tabskip\z@{##}$&\global\@eqcnt\@ne
 \hfil$\displaystyle{{}##{}}$\hfil
 &\global\@eqcnt\tw@$\displaystyle\tabskip\z@{##}$\hfil
 \tabskip\@centering&\llap{##}\tabskip\z@\cr}
\makeatother

\begin{document}

\oddsidemargin 5mm

\begin{titlepage}
\setcounter{page}{0}
\begin{flushright}
EPHOU 96-002\\
April 1996
\end{flushright}

\vs{6}
\begin{center}
{\Large Integral Equations of Fields 
 on the Rotating Black Hole}

\vs{6}
{\large
Takahiro
Masuda
\footnote{e-mail address: masuda@phys.hokudai.ac.jp}
\\ and \\
Hisao Suzuki\footnote{e-mail address: hsuzuki@phys.hokudai.ac.jp}}\\
\vs{6}
{\em Department of Physics, \\
Hokkaido
University \\  Sapporo, Hokkaido 060 Japan} \\
\end{center}
\vs{6}

\centerline{{\bf{Abstract}}}

 It is known that the radial equation of the massless fields with
 spin around Kerr black holes 
cannot be solved by special functions.
 Recently, the analytic solution was obtained by use of 
the expansion in terms of
the special functions and various astrophysical application 
have been discussed. 
It was pointed out that the coefficients of the
expansion by the confluent hypergeometric functions are identical to those
of the expansion by the hypergeometric functions. We explain the
reason of this fact by using the integral equations of the radial equation.
It is shown that the kernel of the equation can be written by the 
product of confluent hypergeometric functions. The integral equaton
transforms the  expansion in terms of the confluent hypergeometric
functions  to that of the hypergeometric functions and vice versa, 
which explains the
reason why the  expansion coefficients are universal.

\end{titlepage}
\newpage

\renewcommand{\thefootnote}{\arabic{footnote}}
\setcounter{footnote}{0}

\sect{Introduction}
\indent

One of the remarkable feature of the Kerr black hole, which is known
to be unique black hole solution with no electro-magnetic 
charge\cite{Robinson,Mazur},
is that the gauge invariant perturbation of the spin $0$\cite{Carter},
the $ 1/2$\cite{Teukolsky} fields,
the electromagnetic fields\cite{Teukolsky}, 
the gravitino\cite{Guven} and the gravity\cite{Teukolsky} can be treated by
separable  equations called Teukolsky equations\cite{Teukolsky}. 
It should be noted that  Whiting\cite{Whiting} provided an analytic 
proof of the stability of
the Kerr black holes by using these equations.
Unfortunately, Teukolsky equation cannot be solved by any popular
special functions since the equation has two regular singular
points at horizons and one irregular singular point at infinity.
In order to determine the physical quantities such as 
the absorption probabilities of black holes,
we usually had to use numerical integration of the equation. However, for the
purpose of investigating the property of the scattering data, it
seems useful to provide analytic solutions as has been achieved
for Coulomb scattering. These may be used as starting points of the
perturbation for the quantum corrections. 
One of physical application is the absorption coefficients of 
the black holes\cite{Otchik,Mano}.
Recently, the analytic methods have been getting a powerful tool
 even in the field of the gravitational wave astrophysics.
\cite{Sasaki,TagoshiSasaki,SSTT,PoissonSasaki}

The technique for analyzing this type of equation is quite old because
a equation similar to Teukolsky equation appeared when we consider
the wave function in terms of spheroidal coordinates, which is known
to be spheroidal wave equation\cite{Flammer,MeixnerShafke}. 
The solution of the equation can be
expanded either by the hypergeometric functions or by the confluent
hypergeometric functions. Therefore it is natural that similar  expansions 
has been considered
even for Teukolsky equation\cite{Leaver,Otchik,Mano}. 

Regarding the structure of the expansion coefficients of Teukolsky equation, 
an interesting observation was given in Refs.\cite{Otchik,Mano} who showed 
that the expansion coefficients in terms of the hypergeometric functions are
identical to those  by the confluent hypergeometric functions
up to re-definition of the coefficients.  
This universality of the expansion coefficients appeared even in the case of 
 the spheroidal wave functions \cite{Flammer,MeixnerShafke}. The wave function
 of three dimensional flat space is separable in the spheroidal coordinates. 
The radial equation and the angular 
equation can be identified by some change of variables, which can be 
identified as a single spheroidal wave equation. The angular equation 
is usually expanded by Legendre functions whereas the radial equation is 
expanded by Bessel functions. It turns out that the expansion coefficients 
are identical up to 
some normalization factor. The reason was explained by use of the fact that 
the angle
equation and the radial equation of spheroidal wave equation are connected
by an integral equation, because of which the coefficients are
identical up to normalization factor\cite{Flammer}.  
This fact indicates that there might be some
transformation which connects
these expansions even for Teukolsky equation. However we cannot obtain the 
integral equation by a simple generalization of the spheroidal functions 
because the construction of the kernel in the case of spheroidal functions 
crusially depends on 
the fact that the original space is just a three dimensional flat space.  
In the case of Teukolsky equation, we cannot 
construct any simple system even if we associate any angle variables. 
Therefore, the above construction cannot be applied.
Instead of considering an analog of such spheroidal wave functions, we treat 
the equation as an analog of Heun's equation\cite{Bateman}. 
Heun's equation\cite{Bateman} has 
four defenite singular points in the equation and Teukolsky equation can 
be considered as a confluent limit of Heun's equation\cite{Ince}. 
We will use a principle used for the   construction of the integral equation 
of the Heun's equation\cite{Erdelyi}. By constructing the 
integral equation for the Teukolsky equation, which can be regarded as a 
confluent analog of the kernel of the Heun's' equation, we will show 
that the expansion coefficients are universal.

In the next section, we review the analytic expansion of the Teukolsky equation. 
There are basically three type of expansions which cannot be connected by the 
analytic continuation.

In section 3, we will construct a integral equation of the Teukolsky equation. It will be shown that the integral kernel can be written by a product of the confluent hypergeometric functions.

In section 4, it will be shown that the these expansions can be connected by the integral transformation, which implies that the expansion coefficients are universal.

Section 5 is devoted to some discussions.
\indent 
\sect{Analytic expansions of Teukolsky equation}
\indent
In the Boyer-Lindquist coordinates and in units such that $c=G=1$, 
the Kerr metric is written as 
\begin{eqnarray}
ds^2=\left(1-{2Mr\over \Sigma}\right)dt^2&+&\left({4Mar\sin^2\theta\over 
\Sigma}\right)dtd\phi-\left({\Sigma\over \triangle}\right)dr^2\nonumber \\
& &-\Sigma d\theta^2-\sin^2\theta\left(r^2+a^2+{2Ma^2r\sin^2\theta\over 
\Sigma}\right)d\phi^2,
\end{eqnarray}
where $M$ is the mass of the black hole, $aM$ its angular momentum, 
$\Sigma=r^2+a^2\cos^2\theta$, and $\triangle=r^2+a^2-2Mr$.
Klein-Gordon equation for massless fields 
$\psi$ in a Kerr black hole background
can be separated by setting $\psi=e^{-i\omega t}e^{im\phi}S^m_l(\theta)
R(r)$, and radial equation is given by
\begin{eqnarray}
\triangle^{-s}{d\over dr}\left(\triangle^{s+1}{dR\over dr}\right)
+\left({K^2-2is(r-M)K\over \triangle}+4is\omega r-\la\right)R=0,
\end{eqnarray}
where $s$ is a parameter called spin weight of the field, 
and $\la=E-2am\omega+a^2\omega^2-s(s+1),$
and $E$ is eigenvalue of spheroidal harmonics $S^m_l(\theta)$. 
 This equation has two regular singularities at $r=r_{\pm}=M\pm\sqrt{M^2-a^2}
=M\pm p$ and an irregular singularity at $r=\infty$. Setting the new variable  
$z=(r_+-r)/2p,$ radial
 equation becomes\cite{Teukolsky}
\begin{eqnarray}
z^2(z-1)^2{\pa^2R\over \pa z^2}&+&(s+1)z(z-1)(2z-1){\pa R\over \pa z}
\nonumber \\
&+&
\left[{K^2\over 4p^2}+{isK(2z-1)\over 2p}-
z(z-1)(8is\omega pz-4is\omega r_+ +\la)
\right]R=0,\label{eq:rad}
\end{eqnarray}
where $s$ takes integer or half integer value and 
$K=(4p^2z^2-4pr_+z+2Mr_+)\omega-am.$ The analytic solution of this equation 
is obtained not by any special functions but by expansions in terms of 
special functions such as the hypergeometric functions and the confluent 
hypergeometric functions.   
 
If we use the expansion in terms of the hypergeometric functions, we have two 
independent solutions
of equation (\ref{eq:rad}). Introducing the slightly modified 
hypergeometric function $P(a,b;c;z)$ 
\begin{eqnarray}
P(a,b;c;z)={\G(a)\G(b)\over \G(c)}F(a,b;c;z),
\end{eqnarray}
where $F(a,b;c;z)$ is the  hypergeometric function, 
we can express two solutions around $z=0$ in the form\cite{Mano}
\begin{eqnarray}
R_1^{\nu}&=&e^{i\sigma z}(-z)^{\rho}(1-z)^{\delta}\sum_{k=-
\infty}^{\infty}h^{1\nu}_kP(a_k,b_k;c;z),
\nonumber \\
R_2^{\nu}&=&e^{i\sigma z}(-z)^{\rho+1-c}(1-z)^{\delta}\sum_{k=-
\infty}^{\infty}h^{2\nu}_kP(a_k-c+1,b_k-c+1;2-c;z),\label{eq:hyp}
\end{eqnarray}
where \begin{equation}
\rho=-s-i\omega M-i\tau,\ \ \delta=i\omega M-i\tau,
\ \ \sigma=2p\omega,\label{eq:fac}
\end{equation}
and $\tau=(2\omega M^2-am)/2p$, and
\begin{equation}
a_k=\rho+\delta+s-k-\nu,\ \
b_k=\rho+\delta+s+k+\nu+1,\ \
c=2\rho+s+1.
\end{equation}
 By inserting the expression $(2.5)$ into $(2.3)$, we find that the 
 expansion coefficients $h^{1\nu}_k$and $h^{2\nu}_k$   
should satisfy the same recursion relations;
\begin{equation}
\alpha_kh^{\nu}_{k+1}+\beta_kh^{\nu}_k+\gamma_kh^{\nu}_{k-1}=0,\label{eq:rec1}
\end{equation}
\begin{eqnarray}
&\alpha_k&={i\sigma (s-2i\omega M+k+\nu+1)(-2i\tau+k+\nu+1)(s+2i\omega M+
k+\nu+1)
\over (2k+2\nu+3)(k+\nu+1)},\nonumber\\
&\beta_k&=8\omega^2M^2-a^2\omega^2-E+(k+\nu)(k+\nu+1)
+{2\sigma\tau(2\omega M+is)
(2\omega M-is)\over (k+\nu)(k+\nu+1)}, \\
&\gamma_k&={i\sigma (s-2i\omega M-k-\nu)(-2i\tau-k-\nu)(s+2i\omega M-k-\nu)
 \over (2k+2\nu-1)(k+\nu)}.\nonumber
\end{eqnarray}
The parameter $\nu$ can be obtained by the convergence of the recursion relation\cite{Otchik,Mano} or by using the post-newtonian principle\cite{Mano}.

Note that the solutions around  $z=1$, which represents the inner horizon, 
can be obtained by the analytic continuation of the hypergeometric functions.
 Two independent solutions around $z=1$ are
\begin{eqnarray}
R^{\nu}_3&=&e^{i\sigma z}(-z)^{\rho}(1-z)^{\delta}\sum_{k=-
\infty}^{\infty}{h^{3\nu}_k\over \G(c-a_k)\G(c-b_k)}P(a_k,b_k;a_k+b_k-c+1;
1-z), \nonumber \\
R^{\nu}_4&=&e^{i\sigma z}(-z)^{\rho}(1-z)^{\delta} \\
&{}&\qquad \times \sum_{k=-
\infty}^{\infty}{h^{4\nu}_k(1-z)^{c-a_k-b_k}\over \G(c-a_k)\G(c-b_k)}
P(c-a_k,c-b_k;c-a_k-b_k+1;1-z). \nonumber 
\end{eqnarray}
Since these expansions are expressed as linear combinations 
of $R^{\nu}_1$ and $R^{\nu}_2$ due to the analytic 
continuation, the recursion relations for 
 $h^{3\nu}_k$ and $h^{4\nu}_k$ are identical to those of 
$h^{1\nu}_k,$ and $h^{2\nu}_k$.

Other type of solutions are given by the expansion 
in terms of the confluent hypergeometric functions.
 Introducing the new variable $\rho=-2\omega pz,$ 
one type of the expansions is given by \cite{Otchik,Mano}
\begin{eqnarray}
R_1^{\nu}&=&\rho^{\delta}(\rho+2\omega p)^{-\delta-s}
\sum_{k=-\infty}^{\infty}g^{1\nu}_kG_{k+\nu}(\eta,\rho),
\nonumber \\
R_2^{\nu}&=&\rho^{\delta}(\rho+2\omega p)^{-\delta-s}
\sum_{k=-\infty}^{\infty}g^{2\nu}_kG_{-k-\nu-1}(\eta,\rho),\label{eq:conf}
\end{eqnarray}
where $\eta=-(2\omega M+is)$ and $\delta
=i\omega M-i\tau,$ and $G_{l}(\eta,\rho)$ 
satisfies the Coulomb type equation
\begin{eqnarray}
\rho{\pa^2\over \pa \rho^2}G_{l}(\eta,\rho)+2{\pa\over \pa\rho}G_{l}
(\eta,\rho)+
\left(\rho-2\eta-{l(l+1)\over \rho}\right)G_{l}(\eta,\rho)=0,
\end{eqnarray}
which is expressed in terms of Kummer's confluent hypergeometric function 
$\Phi(a;b;\rho)$ as 
\begin{eqnarray}
G_{l}(\eta,\rho)={\Gamma(l+1+i\eta)\over e^{\pi\eta/2}\Gamma(2l+2)}
e^{-i\rho}(2i\rho)^{l}\Phi(l+1-i\eta;
2l+2;2i\rho).
\end{eqnarray}
By inserting (\ref{eq:conf}) into the original equation, we find that  
the expansion coefficients $g^{1\nu}_k$ and $g^{2\nu}$ should satisfy 
the following recursion relations;
\begin{equation}
\alpha'_kg^{\nu}_{k+1}+\beta'_kg^{\nu}_k+\gamma'_kg^{\nu}_{k-1}=0,
\end{equation}
\begin{eqnarray}
&\alpha'_k&=\alpha_k, \nonumber \\
&\beta'_k&=\beta_k,\\
&\gamma'_k&=\gamma_k, \nonumber
\end{eqnarray}
which is identical to (\ref{eq:rec1})

Moreover, if we set $\chi=2\omega p(1-z)$, which is the expansion 
in terms of the  inner horizon, another expansion can be 
obtained in the form \cite{Leaver}
\begin{eqnarray}
R^{\nu}_3&=&\chi^{\e}(\chi-2\omega p)^{-s-\e}\sum_{k=-\infty}^{\infty}
{g^{3\nu}_{k}\over \G(c-a_k)\G(c-b_k)}G_{k+\nu}(\eta,\chi),\nonumber \\
R^{\nu}_4&=&\chi^{\e}(\chi-2\omega p)^{-s-\e}\sum_{k=-\infty}^{\infty}
{g^{4\nu}_k\over \G(c-a_k)\G(c-b_k)}G_{-k-\nu-1}(\eta,\chi),\label{eq:conflu}
\end{eqnarray}
where $\e=-s-i\omega M-i\tau$. 
It turns out that the recursion relation for $g^{3\nu}_k$ and $g^{4\nu}_k$ 
are again identical to those of $g^{1\nu}_k$ and $g^{2\nu}_k$ and $h^{\nu}_k$. 
In this case,
 we cannot explain the reason of having identical recursion relations 
simply from the analytic continuation on the contrary to the case of 
hypergeometric functions.  
 
In the above construction, the coefficients $g^{\nu}_k$ and $h^{\nu}_k$ 
are identical 
up to normalization factor as
\begin{equation}
g^{\nu}_k\sim  h^\nu_k \label{eq:expand}.
\end{equation}
 Since $h^{\nu}_k,\ g^{\nu}_k$ are expansion coefficients 
in terms of different kinds of special functions, such as the
hypergeometric function and the confluent hypergeometric function, which 
have different analytic properties and different 
asymptotic behaviors, it is not trivial why the relation (\ref{eq:expand}) 
holds. Moreover, the identification of the coefficients $g_k^{1\nu}$ and 
$g_k^{3\nu}$ 
is also the subject which cannot be explained by any analytic continuation 
of the special functions. In the next section, we are going to  
explain the reason by use of the integral equations.
  
\indent 
\sect{Integral equation}
\indent
We are going to construct the integral equation in terms of 
$R(z)$ which satisfies the equation (\ref{eq:rad});
\begin{eqnarray}
&M_z& R(z) \\ \nonumber
&\equiv&z(z-1)\left\{ {\pa^2R(z)\over \pa z^2}+(s+1)\left({1\over z}+
{1\over z-1}\right){\pa R(z)\over \pa z}\right\}
\nonumber \\
&{}&\quad+\left[{K^2\over 4p^2z(z-1)}+{isK(2z-1)\over 2pz(z-1)}-
8is\omega pz+4is\omega r_+ -\la
\right]R(z)=0.
\end{eqnarray}
The integral transformation which maps one solution to other solution is
 given by 
\begin{eqnarray}
R'(x)=\int_{\cal C}y^s(y-1)^sK(x,y)R(y)\,dy,
\end{eqnarray}
where the function $R'(x)$ is also a solution of equation (\ref{eq:rad}) 
if the kernel satisfy the condition
\begin{equation}
(M_x-M_y)K(x,y)=0,\label{eq:kernel}
\end{equation}
and the surface term of the integral
\begin{eqnarray}
y^{s+1}(y-1)^{s+1}\left\{ {\pa K(x,y)\over \pa y}R(y)-K(x,y){
\pa R(y)\over \pa y}\right\}
\end{eqnarray} 
vanishes at the end of $\cal C$\cite{Ince}. To find the kernel, we set  
new variables as 
\begin{eqnarray}
\xi=xy,\ \ \ \ \zeta=(x-1)(y-1).\label{eq:variable}
\end{eqnarray}
Then the equation (\ref{eq:kernel}) becomes  
\begin{eqnarray}
\xi{\pa^2K\over \pa \xi^2}&+&(s+1){\pa K\over \pa \xi} \nonumber \\ 
&{}&\qquad +\left\{ 4p^2\omega^2\xi-8pM\omega^2-4is\omega p+
{(M\omega +\tau)(M\omega+\tau-is)\over \xi}\right\}K \\ \nonumber
&-&\zeta{\pa^2K\over \pa \zeta^2}-(s+1){\pa K\over \pa \zeta}
-\left\{ 4p^2\omega^2\zeta+{(M\omega-\tau)(M\omega -\tau-is)\over 
\zeta}\right\}K=0.
\end{eqnarray}
Therefore we can separate the variable, and we put 
$K(x,y)=P(\xi)Q(\zeta)$ so that $P(\xi)$ and $Q(\zeta)$ 
satisfy the equations
\begin{eqnarray}
\xi{\pa^2P(\xi)\over \pa \xi^2}&+&(s+1){\pa P(\xi)\over \pa \xi}\\ \nonumber
&{}&+\left\{ 4p^2\omega^2\xi-8pM\omega^2-4is\omega p+
{(M\omega +\tau)(M\omega+\tau-is)\over \xi}\right\}P(\xi)=\la P(\xi), \\
\zeta{\pa^2Q(\zeta)\over \pa \zeta^2}&+&(s+1){\pa Q(\zeta)\over \pa \zeta}
+\left\{ 4p^2\omega^2\zeta+{(M\omega-\tau)(M\omega -\tau-is)\over
\zeta}\right\}Q(\zeta)=\la Q(\zeta),
\end{eqnarray}
where $\la$ is a separation constant.
Each equation has two independent solutions and these are expressed by 
using Kummer's confuent hypergeometric functions as
\begin{eqnarray}
P_1(\xi)&=&e^{i\sigma \xi}\xi^{\rho}\Phi\left({2\rho+s+1\over 2}-{E+\la\over 
2i\sigma};2\rho+s+1;-2i\sigma \xi\right), \nonumber \\ 
P_2(\xi)&=&e^{i\sigma\xi}\xi^{-\rho-s}\Phi\left(
-{2\rho+s-1\over 2}-{E+\la\over 2i\sigma};1-2\rho-s;-2i\sigma \xi
\right),\\
Q_1(\zeta)&=&e^{i\sigma \zeta}\zeta^{\delta}\Phi\left(
{2\delta+s+1\over 2}-{\la\over 2i\sigma};2\delta++s+1;-2i\sigma \zeta\right)
,\nonumber \\ 
Q_2(\zeta)&=&e^{i\sigma \zeta}\zeta^{-\delta-s}\Phi\left(
-{2\delta+s-1\over 2}-{\la\over 2i\sigma};1-2\delta-s;-2i\sigma \zeta\right),
\label{eq:kersol}
\end{eqnarray}
where $\rho=-s-iM\omega-i\tau,\ \delta=iM\omega-i\tau,\ \sigma=2p\omega$ and 
$E=8M\omega^2 p+4is\omega p$.
We can thus solve the kernel equation in a general form. 
Namely, the kernel $K(x,y)$ is expressed as the simple 
product of these solutions, or as the product of 
some linear combinations of these solutions such as Whittaker's function.
 Since $\la$ is a arbitrary constant, we choose $\la$ in such 
a way that the kernel  
becomes a simple form in order to evaluate the integral transformation 
easily. The integral equation maps a solution of Teukolsky equation to 
another solution. Therefore, we can transform a analytic expansion to 
other expansions by various choice of the kernel. 
  In the next section, we are going  to perform the integral transformation of 
such various expansions of equation (\ref{eq:rad}).

\indent
\sect{Integral transformations}
\indent

First of all, we consider the transformation from the 
hypergeometric expansions (\ref{eq:hyp}) to Coulomb expansions 
 (\ref{eq:conf}) around $z=0$; 
we take $R_1^{\nu}(y)$ in (\ref{eq:hyp}) as $R(y)$
\begin{eqnarray}
R_1^{\nu}=e^{i\sigma z}(-y)^{\rho}(1-y)^{\delta}\sum_{k=-
\infty}^{\infty}h^{1\nu}_kP(a_k,b_k;c;y). 
\end{eqnarray}
A suitable choice of the kernel for this transformation 
is as follows. We choose $P(\xi)$ in the kernel  
 by using Whittaker's function $M_{\kappa,\mu}(z)$ as 
\begin{eqnarray}
P(\xi)=\xi^{-(s+1)/2}M_{\kappa,\mu}(-2i\sigma \xi),
\end{eqnarray}
where $\kappa=(E+\la)/2i\sigma$ and $\mu=\rho+s/2$. In order to eliminate 
the factor $(1-y)$ in the integral equation, we take $Q_2(\zeta)$ in (
\ref{eq:kersol}) for  
$Q(\zeta)$. For the convergence of the integral, we take $\la=
i\sigma(-2\delta-s+1)$ so that $Q(\zeta)$ becomes
\begin{eqnarray}
Q(\zeta)=e^{i\sigma \zeta}
\zeta^{-\delta-s}.
\end{eqnarray}
Then the integral equation is given by 
\begin{eqnarray}
R'_1(x)&=&(const.)x^{-(s+1)/2}(x-1)^{-\delta-s}e^{-i\sigma(x-1)}\nonumber \\ 
&\times& \int_{\cal C}e^{i\sigma xy}y^{-(s+1)/2+\rho+s}M_{\kappa,\mu}(
-2i\sigma xy)\sum_{k=-\infty}^{\infty}h^{1\nu}_kP(a_k,b_k;c;y)\,dy.
\end{eqnarray}
 By taking the region of the integration  as the
 interval from $0$ to $+\infty$ and using the formulae in \cite{int}, we can evaluate the integral as  
\begin{eqnarray}
R'_1(\rho)=(const.)
\rho^{\delta}(\rho&+&2\omega p)^{-\delta-s}\sum_{k=-\infty}^{\infty}
h^{1\nu}_k \nonumber \\ 
&\times&\Bigl\{ C_1\G(2k+2\nu+1)\G(-2k-2\nu)G_{-k-\nu-1}(\eta,\rho) 
\nonumber \\ 
&{}&\quad+C_2\G(-2k-2\nu-1)\G(2k+2\nu+2)G_{k+\nu}(\eta,\rho)\Bigr\},
\label{eq:trans1}
\end{eqnarray}
where $C_1,\ C_2$ are constants which are independent of $k$. Note that 
if we start with $R^{\nu}_2$, we obtain the same result 
as the equation (\ref{eq:trans1}) after the integral 
transformation by using an appropriate kernel.  

Note that the solution $R'_1(\rho)$ consists of two independent solutions which
are expanded in terms of Coulomb type functions and they are
identical to the solution (\ref{eq:conf}). Thus 
two kinds of expansions are connected by the
integral transformation. Recognizing the coefficients
 of $G_{\nu+k}(\eta,\rho),\ G_{-k-\nu-1}(\eta,\rho)$ as $g^{1\nu}_k,\ 
g^{2\nu}_k$ respectively, the relation between $h^{1\nu}_k$
and $g^{1\nu}_k,\ g^{2\nu}_k$ is 
\begin{eqnarray}
g^{1\nu}_k&\sim&\G(2k+2\nu+1)\G(-2k-2\nu)h^{1\nu}_k\sim h^{1\nu}_k,\\ 
g^{2\nu}_k&\sim&\G(-2k-2\nu-1)\G(2k+2\nu+2)h^{1\nu}_k\sim h^{1\nu}_k,
\end{eqnarray}
Therefore the fact that 
coefficients $h^{\nu}_k,\ g^{\nu}_k$ of different kinds of expansions 
satisfy the same recursion relations can be  understood quite naturally.

Let us consider the inverse transformation.
In this case, we start with a slightly modified form
 of $R_1^{\nu}(\rho)$ in (\ref{eq:conf}) as $R(y)$ 
\begin{eqnarray}
R_1^{\nu}(\rho)&=&\rho^{\delta}(\rho+2\omega p)^{-\delta-s}\sum_{k=-
\infty}^{\infty}g^{1\nu}_kG_{k+\nu}(\eta,\rho),\nonumber \\
&=&\rho^{\delta}(\rho+2\omega p)^{-\delta-s}\sum_{k=-
\infty}^{\infty}g^{1\nu}_k{\G(k+\nu+1+i\eta)(2i)^{k+\nu+1}\over 
e^{\pi\eta\over 2}\G(2k+2\nu+2)\rho}M_{i\eta,k+\nu+1/2}(2i\rho).
\end{eqnarray}
Our choice of the kernel is as follows. In order to eliminate the factor
 $(y-1)$, we take $Q_1(\zeta)$ with $\la=-i\sigma(2\delta+s+1)$ so that
\begin{eqnarray}
Q(\zeta)=e^{-i\sigma\zeta}\zeta^{\delta}.
\end{eqnarray} 
For the convergence of the integral, 
we combine $P_1(\xi),P_2(\xi)$ into 
the form
\begin{eqnarray}
P(\xi)=\xi^{-(s+1)/2}W_{-\kappa,\mu}(2i\sigma \xi),
\end{eqnarray}
where $W_{-\kappa,\mu}(z)$ is Whittaker's function.
Then the integral equation becomes
\begin{eqnarray}
R'_1(x)&=&(const.)x^{-(s+1)/2}(x-1)^{\delta}e^{i\sigma(x-1)} \nonumber \\ 
&\times&\int_{\cal C}
y^{s-(s+1)/2+\delta-1}e^{-i\sigma(x-1)y}
W_{-\kappa,\mu}(2i\sigma xy)\\
&{}&\qquad \times\sum_{k=-\infty}^{\infty}
g^{1\nu}_k{\G(k+\nu+1+i\eta)\over \G(2k+2\nu+2)}(2i)^{k+\nu+1}
M_{i\eta,k+\nu+1/2}(-2i\sigma y)\,dy. \nonumber
\end{eqnarray}
By taking the region of the integration as the interval 0 to infinity and using the formulae in \cite{table}, we can evaluate the integral.
 The solution is 
\begin{eqnarray}
R'_1(x)=(const.)x^{\rho}(1-x)^{\delta}&e^{i\sigma x}&\sum_{k=-\infty}^{\infty}
g^{1\nu}_k \left\{C_1P(a_k,b_k;c;x)\right.
\nonumber \\
 &+&C_2\left.x^{1-c}P(a_k-c+1,b_k-c+1;2-c;x) \right\}.
\end{eqnarray}
Thus we could perform the inverse transformaion, which completes the 
relation between two expansion around $z=0$. The relation 
between expansion coefficients are
\begin{eqnarray}
h^{1\nu}_k\sim h^{2\nu}_k\sim g^{1\nu}_k\sim g^{2\nu}_k.
\end{eqnarray}

 We next consider the relation between expansions around $z=1$. 
As before, we use $R^{\nu}_3(\chi)$ in (\ref{eq:conflu}) 
as $R(y)$ in the modified form 
\begin{eqnarray}
R^{\nu}_3(\chi)&=&\chi^{\e}(\chi-2\omega p)^{-\e-s}
\sum_{k=-\infty}^{\infty}{g^{3\nu}_k\over \G(c-a_k)\G(c-b_k)}
G_{k+\nu}(\eta,\chi),\nonumber \\
&=&\chi^{\e-1}(\chi-2\omega p)^{-\e-s}\\
&{}&\qquad \times \sum_{k=-\infty}^{\infty}{g^{3\nu}_k\G(k+\nu+1+i\eta)
(2i\sigma)^{k+\nu+1}
\over \G(c-a_k)\G(c-b_k)e^{\pi\eta\over 2}\G(2k+2\nu+2)}
M_{i\eta,k+\nu+1/2}(2i\chi),\nonumber
\end{eqnarray}
where $\chi=2\omega p(1-y)$ and $\e=-s-i\omega M-i\tau$. 
In this case, the role of $y$ 
and $1-y$ are interchanged in the integral equation. 
In order to eliminate the 
factor $y$,  we take $P(\xi)$ as
\begin{eqnarray}
P(\xi)=\xi^{\e}e^{i\sigma \xi},
\end{eqnarray}
where we set $\la=i\sigma(2\e+s+1)-E$. For the convergence of the integral, 
we take $Q(\zeta)$ as
\begin{eqnarray}
Q(\zeta)=\zeta^{-(s+1)/2}W_{\kappa',\mu'}(-2i\sigma \zeta),
\end{eqnarray}
where $\kappa'=\la/2i\sigma,\ \mu'=\delta+s/2$.
Then the integral transformation becomes
\begin{eqnarray}
R'^{\nu}_3&=&(const.)x^{\e}(1-x)^{-(s+1)/2}e^{i\sigma x}\nonumber \\
&\times&\int_{\cal C}(1-y)^{s-(s+1)/2+\e-1}e^{i\sigma x(y-1)}
W_{\kappa',\mu'}(-2i\sigma(1-x)(1-y))\\
&{}&\times \sum_{k=-\infty}^{\infty}{g^{3\nu}_k\G(k+\nu+1+i\eta)
(2i\sigma)^{k+\nu}\over \G(c-a_k)\G(c-b_k)\G(2k+2\nu+2)}M_{i\eta,k+\nu+1/2}
(2i\sigma(1-y))\,dy.\nonumber
\end{eqnarray}
By considering  the integral region as the interval 0 to infinity, and by using the formulae in \cite{table}, we obtain the solution of the equation as  
\begin{eqnarray}
R'^{\nu}_3&=&(const.)x^{\e}(1-x)^{\delta}e^{i\sigma x}
\sum_{k=-\infty}^{\infty}g^{3\nu}_k\nonumber \\
&{}&\left\{ {C_1\over \G(c-a_k)\G(c-b_k)}P(a_k,b_k;a_k+b_k-c+1;1-x)\right. \\
&{}&+\left.{C_2
(1-x)^{c-a_k-b_k}\over \G(c-a_k)\G(c-b_k)}
P(c-a_k,c-b_k;c-a_k-b_k+1;1-x)\right\}.\nonumber
\end{eqnarray}
Note that the same result holds if we start with $R^{\nu}_4(\chi)$.

$R'^{\nu}_3$ consists of two independent solution which are expanded in 
terms of hypergeometric functions around $z=1$. Thus two kinds of expansion
 around $z=1$ are connected by the integral transformation. 
The relation between expansion coefficients are
\begin{eqnarray}
g^{3\nu}_k\sim g^{4\nu}_k\sim h^{3\nu}_k\sim h^{4\nu}_k.
\end{eqnarray}
We thus find that all the coefficients are connected by the analytic 
continuation of the wave function and the integral transformation which 
maps the solution of Teukolsky equation to other solutions.

\indent
\section{Conclusion}
\indent

We have constructed the integral equation in terms of equation (\ref{eq:rad}), 
and solve the kernel in the general form as the product of special functions. 
 By various choice of the kernel, 
we have performed  the integral transformaions which connect various 
expansions of equation (\ref{eq:rad}).  In all regions of 
$z$, we can relate these expansions one another 
by the integral transformation, and by the  
analytic continuation of hypergeometric 
functions. 
As a consequence,
the coefficients $g^{i\nu}_k$ of Coulomb type expansions 
satisfy the same recusion relations of the 
coefficients $h^{i\nu}_k$ of the hypergeometric expansions, and 
they are identical up to 
normalization factors.      

In any case of integral transformations, the kernel is expressed  
as the product of the confluent hypergeometric functions. It is this property 
that makes us possible to perform the integral transformation.
 
Let us consider the case of the spheroidal equation. 
The equation (\ref{eq:kernel}) is 
just the wave equation in the spheroidal coordinates where 
the variable $x$ is the radial coordinate and  $y$ is the 
coordinate representing the angle\cite{Flammer}. 
In other words, $x$ and $y$ can be treated as  dual coordinates. 
It was quite easy to obtain the integral equation of spheroidal wave 
functions  because 
we know other type of separable coordinates in flat space. 
On the other hand, in the case of Teukolsky equation, we have 
dealt with dual coordinates of the 
radial function which are not the spheroidal coordinates in Kerr geometry
 for the construction of integral equation. 
The existence of other separable variable (\ref{eq:variable}) 
seems to show some kind of symmetry of the space-time in a more wider 
geometry. 
Further study on this point of view seems interesting.

\vskip 2cm
{\Large Acknowledgment}

One of the author (H. S) would like to thank E. Takasugi and S. Mano for discussions. 



\newpage
\begin{thebibliography}{99}
\bibitem{Robinson}
D.C.Robinson,
\PRL{34}, 905 (1975).
\bibitem{Mazur}
P.O.Mazur,
Gen. Rel. and Grav. 16, 211 (1984).
\bibitem{Carter}
B.Carter,
\CMP{10}, 280 (1968).
\bibitem{Teukolsky}
S.A.Teukolsky,
\AJ{185}, 635 (1973).
\bibitem{Guven}
R.G\"{u}ven,
\PR{D 22}, 2327 (1980).
\bibitem{Whiting}
B.F.Whiting,
\JMP{30},1301 (1989).
\bibitem{Otchik}
V.S.Otchik, in \it Quantum Systems; Newtends and Methods, \rm
edited by A.I.Barnt et.al.(World Scientific,1995), p.154.
\bibitem{Mano}
S.Mano, H.Suzuki and E.Takasugi, Osaka University preprint OU-HET 238,
\\ gr-qc/9603020.
\bibitem{Sasaki}
M.Sasaki,
\PTP{92},17 (1994).
\bibitem{TagoshiSasaki}
H.Tagoshi and M.Sasaki, \PTP{92},745 (1994).
\bibitem{SSTT}
M.Shibata, M.Sasaki, H.Tagoshi and T.Tanaka,
Phys. Rev. \bf D 51\rm, 1646 (1995).
\bibitem{PoissonSasaki}
E.Poisson and M.Sasaki, Phys. Rev. \bf D 51\rm, 5753 (1995).
\bibitem{Flammer}
C.Flammer,
\it Spheroidal Wave Functions \rm (Stanford University Press,Calfornia, 1957).
\bibitem{MeixnerShafke}
J.Meixner and F.W.Sh\"afke,
\it Mathieushe Funktionen und Sph\"aroidfunktionen \\ \rm (Berlin, 1954).
\bibitem{Leaver}
E.W.Leaver,
\JMP{27},1238 (1986).
\bibitem{Bateman}
see for example, A.Erd\'{e}lyi et.al., \it Higher
Transcendal Fuctions \rm (McGraw-Hill, New York)
 Vol.\bf 3. \rm
\bibitem{Ince}
E.L.Ince, \it Ordinary Differential Equations \rm (Dover Publications, 1956).
\bibitem{Erdelyi}
A.Erd\'{e}lyi, Quart.J.of Math. \bf 13\rm, 107 (1942).
\bibitem{int}
A.P.Prudnikov,Yu.A.Brychkov and O.I.Marichev,
\it Integrals and Series \\
\rm (Gordon and Breach Science Publishers) Vol.\bf 3, \rm p.352.
\bibitem{table}
A.Erd\'elyi et al., \it Tables of Integral Transforms \rm
(McGraw-Hill, New York, 1954) \\
Vol.\bf 2, \rm p.416.

\end{thebibliography}
\end{document}